\documentclass[aps,epsf,twocolumn,showpacs,floatfix]{revtex4}
\usepackage{amsmath}
\usepackage{epsfig}

\begin{document}

\title{Magnetic-field dependence of transport in normal
and Andreev billiards:\\ a classical interpretation to the
averaged quantum behavior}

\author{Nikolaos G. Fytas}
\author{Fotis K. Diakonos}
\affiliation{
Department of Physics, University of Athens, GR-15771 Athens, Greece
}

\author{Peter Schmelcher}
\affiliation{
Theoretische Chemie, Im Neuenheimer Feld 229,
Universit\"{a}t Heidelberg, 69120 Heidelberg, Germany
}
\affiliation{
Physikalisches Institut, Philosophenweg 12, Universit\"{a}t Heidelberg,
69120 Heidelberg, Germany
}

\author{Matthias Scheid}
\author{Andreas Lassl}
\author{Klaus Richter}
\affiliation{
Institut f\"{u}r Theoretische Physik, Universit\"{a}t
Regensburg, 93040 Regensburg, Germany
}

\author{Giorgos Fagas}
\affiliation{
Tyndall National Institute, Lee Maltings, Prospect Row, Cork, Ireland
}

\date{\today}

\begin{abstract}
We perform a comparative study of the quantum and classical transport probabilities
of low-energy quasiparticles ballistically traversing normal and Andreev
two-dimensional open cavities with a Sinai-billiard shape.
We focus on the dependence of the transport on the strength of an applied
magnetic field $B$.  With increasing field strength
the classical dynamics changes from mixed to regular phase space. 
Averaging out the quantum fluctuations, we find an
excellent agreement between the quantum and classical transport coefficients
in the complete range of field strengths. This allows an overall description of
the non-monotonic behavior of the average magnetoconductance in terms of the
corresponding classical trajectories, thus, establishing a basic tool useful in the
design and analysis of experiments.
\end{abstract}

\pacs{05.60.Cd, 05.60.Gg, 74.45.+c, 73.23.Ad}

\maketitle

\section{Introduction}
\begin{figure}[b]
\includegraphics[width=8cm]{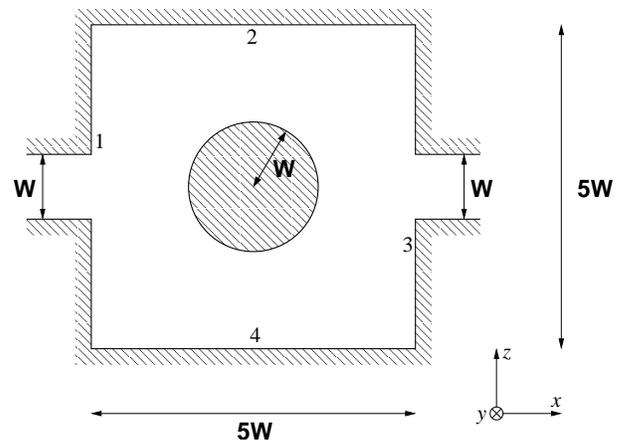}
\caption{\label{fig:fig1} The open geometry of the Sinai billiard
considered in this study.}
\end{figure}

Ballistic transport of particles across billiards
is a field of major importance due to its fundamental properties as well as physical
applications (see for example the reviews~\cite{QuaCh,Alha,Ri00,Jalab}).
In such systems, a two-dimensional cavity is defined by a steplike
single-particle potential where confined particles can propagate
freely between bounces at the billiard walls.
For open systems the possibility of particles being injected and
escaping through holes in the boundary is also allowed.
As an example, we consider the open geometry of the extensively studied
Sinai billiard shown in Fig.~\ref{fig:fig1}.
Experimental realizations are based on exploiting the analogy between
quantum and wave mechanics in either microwave and acoustic cavities
or vibrating plates~\cite{QuaCh}, and on structured two-dimensional electron
gases in artificially tailored semiconductor heterostructures~\cite{Alha,Ri00,Jalab}.
In the latter case, the particles are also charge carriers making these
nanostructures relevant to applied electronics.

\begin{figure}[b]
\includegraphics{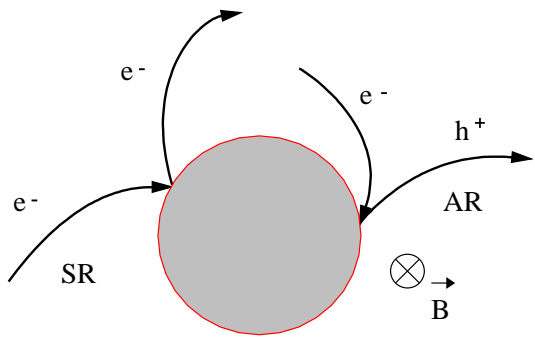}
\caption{\label{fig:fig2} Typical specular (SR) and Andreev reflection (AR) at
the circular central ``antidot'' of Fig.~\ref{fig:fig1}. A
magnetic field is applied as indicated.}
\end{figure}

Focussing the attention on the electronic analogues,
more recently the possibility to couple a superconductor 
to a ballistic quantum dot has been considered
both theoretically~\cite{KosMasGol,S-billiards} and
experimentally~\cite{EPL02ETW}, so that
some part of the billiard boundary exerts the additional property
of Andreev reflection~\cite{And64}.
During this process particles with energies much smaller than the
superconducting gap $\Delta$ are coherently scattered from the
superconducting interface as Fermi sea holes
back to the normal conducting system (and vice versa). Classically, Andreev reflection
manifests itself by retroreflection, i.e., all velocity components are inverted,
compared to the specular reflection where only the boundary normal
component of the velocity is inverted. Thus, Andreev reflected particles
(holes) retrace their trajectories as holes (particles). If, however,
a perpendicular magnetic field is applied in addition, such retracing
no longer occurs due to the inversion of both the charge and the effective
mass of the quasiparticle resulting in opposite bending.
Typical trajectories are illustrated in
Fig.~\ref{fig:fig2}. Here, we investigate the interplay
between trajectory bending and Andreev reflection and demonstrate
how such effects influence the overall (magneto)transport properties
of Andreev billiards when compared to their normal counterparts.

A unique feature of this class of 
(quantum) mechanical systems is their 
suitability for studying the quantum-to-classical correspondence.
In particular, much effort has been devoted in revealing
the quantum fingerprints of the classical dynamics which
may be parametrically tuned from regular to chaotic via, e.g., changes
in the billiard-shape. A range of theoretical tools has been used, spanning
the usual analysis of classical trajectories and the semiclassical
approximation to the models of Random Matrix Theory and fully quantum
mechanical calculations. The main signatures of classical
integrability (or lack of it) on the statistics of energy levels and properties
of the transport coefficients for closed and open systems, respectively,
have been discussed in detail in various reviews~\cite{QuaCh,Alha,Ri00,Jalab}.
Discussions on modifications owing to the possibility of Andreev reflection
appear in more recent studies
\cite{KosMasGol,MBFC96,S-billiards,PRB00CBA,PRL99SB,EPJB01ILV,TA01,PRB04CPP,FTPR05},
mostly focusing on the features of the quantum mechanical level density.

In a similar fashion, the aim of this paper is to determine how far
a purely classical analysis may provide qualitative rationalization
and quantitative predictions for the average quantum mechanical transport
properties of a generic billiard such as that of Fig.~\ref{fig:fig1};
both in the presence or absence of Andreev reflection.
Indeed, by performing exact calculations for the classical and quantum dynamics
of low-energy quasiparticles we find that the classical transport
probabilities of electrons and holes, if appropriate, are in good quantitative
agreement with the mean value (to be defined below) of the corresponding quantum
mechanical scattering coefficients that determine the magnetoconductance of
such systems. 
While most of the previous works considered the case of 
zero or small magnetic field (such that the classical dynamics is not altered),
we particularly analyse the regime of finite magnetic field
strengths and show that the classical trajectories which depend
parametrically on the applied magnetic field suffice to describe the overall features
of the observed non-monotonic behavior.

The article is structured as follows.
In Sec.~\ref{sec:S2}, after a brief discussion on the details
of the studied system, we present 
precise numerical results of the
magnetic-field dependence of the transport coefficients
as determined by the quantum mechanical scattering matrix.
In Sec.~\ref{sec:S3} we first discuss the model describing the corresponding 
classical dynamics and provide an analysis for both the normal and the Andreev
version of the Sinai billiard in Sections~\ref{sec:S3a}
and~\ref{sec:S3b}, respectively. A synopsis is given in Sec.~\ref{sec:S4}.

\section{Quantum mechanical transport properties}
\label{sec:S2}
We consider ballistic transport of charge carriers
in the $2D$ Sinai-billiard shown in Fig.~\ref{fig:fig1} under an externally
applied magnetic field. The side length of the square cavity is
taken $L=5 W$ where $W$ is the width of each of the leads attached to the
left and right of the cavity. The latter define source and sinks
of (quasi)particles. The central scattering disk possesses the radius $R=W$, and
it can be either a normal or a superconducting antidot. In the former case
the antidot represents an infinitely high potential barrier while in the 
latter case it is considered as an extended homogeneous superconductor
characterized by the property of Andreev reflection~\cite{KosMasGol}.
Experimentally, such antidot structures have been realized
in periodic arrangements, thus forming superlattices~\cite{Ri00,EPL02ETW}.
The boundaries of the square cavity, numbered clockwise by the labels 
1 through 4, are always normal conducting potential walls of infinite height.

In the presence of a superconductor the quantum dynamics of the system can be
described by the Bogoliubov-deGennes Hamiltonian:
\begin{equation}
  \label{eqn:BdG_Hamiltonian}
  \mathcal{\hat H} = \left( \begin{array}{cc} \hat H_0 & \hat \Delta  \\
                      \hat \Delta^* & -\hat H_0^* \end{array}\right),
\end{equation}
where the diagonal operators determine the motion of
particles and holes, respectively, and the off-diagonal elements take care
of the coupling between particle- and hole-like excitations.
Later, in our classical calculations we assume perfect Andreev reflection
meaning that {\em all} particles that hit the normal-superconducting (NS) interface are
{\em exactly} retroreflected. In order to model this quantum mechanically
we have to consider perfect coupling between the normal-conducting region
and the superconductor and simulate a bulk superconductor. To this end,
we take its size to be much larger than the superconducting coherence length
$\xi_S=\hbar v_F/2\Delta$; $v_F$ is the Fermi velocity.
Under these conditions, it is sufficient to consider a stepfunction-like behavior
of the pair potential so that $\Delta=\Delta_0$ is constant inside the superconducting
region and zero outside. We also assume that the temperature is
sufficiently smaller than the superconducting critical temperature so that $\xi_S$
does not diverge.

For our numerical calculations we use a discretized version of the
Bogoliubov-deGennes Hamiltonian~(\ref{eqn:BdG_Hamiltonian}) resembling the
tight-binding approximation on a square lattice~\cite{FTPR05}. Hence,
$\hat{ \mathcal{H}}$ becomes a matrix where only coupling between neighboring
lattice sites is considered. The submatrix $[H_0]_{ij}$ has elements
$\epsilon_i-E_F$ for $i=j$ and $\gamma_{ij}$ for nearest neighbors $i$ and $j$.
The Fermi energy $E_F$ is set to a value that allows six open channels
in the leads. The pairing matrix is given by
$[\Delta]_{ij} = \Delta_0\delta_{ij}$ if lattice point $i$ is inside the
superconducting region and it is zero otherwise.
To reproduce the correct dispersion relation in the continuum limit the
onsite energies $\epsilon_i$ and the hopping energies $\gamma_{ij}$
have to fulfill the relation $\epsilon_i =
\sum_{\langle i,j\rangle}\gamma_{ij}$, where $\langle i,j\rangle$ denotes a
summation over nearest neighbors $j$ of site $i$,
see e.g. Ref.~\onlinecite{Fer97}.

In the presence of a magnetic field the hopping energies acquire a phase according
to the Peierls substitution~\cite{Fer97}, $\gamma_{ij}=-\exp[2\pi i
\vec{A}\cdot(\vec{r}_i-\vec{r}_j)/\Phi_0]$. Here, $\vec{A}$ is the
vector potential, $\vec{r}_i-\vec{r}_j$ is the vector
pointing from site $j$ to site $i$ and $\Phi_0=h/e$ is the flux quantum. In
general, the pair potential $\Delta_0\exp{[i\chi(\vec{A})]}$ is also a complex
number. However, it can be chosen real ($\chi\equiv 0$) if the vector potential
$\vec{A}$ is parallel to the screening currents near the NS interface~\cite{Ket99}.
This is achieved by choosing the symmetric gauge
$\vec{A}=[(B/2)z, 0, -(B/2)x]$ that accounts for a
homogeneous magnetic field of strength $B$ in $y$ direction,
perpendicular to the two-dimensional system. In what follows, we define
as magnetic field unit the value $B_{0}=mv_F/(-q_e W)$ for which the cyclotron
radius is equal to $W$.

The transport coefficients are calculated via a recursive decimation method, as
explained in Ref.~\onlinecite{Tad99}. This method enables the exact computation
of the full scattering matrix $S_{n,n^\prime}(\varepsilon,\hat H)$,
which yields scattering properties of quasiparticles with energy $\varepsilon$,
incident on a phase-coherent structure described by a Hamiltonian $\hat H(\vec B)$.
$\left| S_{n,n^\prime}(\varepsilon,\hat H) \right|^2$
is the outgoing flux of quasiparticles along channel $n$, arising from a
unit incident flux along channel $n^\prime$. The quantum numbers $n$ indicating
open scattering channels are conveniently written as $n=(i,\alpha,\nu)$, where $i$
indicates the leads, $\alpha$ takes the discrete values $e$ and $h$ for
particles and holes, respectively, and $\nu$ labels the quantum numbers associated
with the quantization of the wavefunction in the transverse direction.
As shown in Ref.~\onlinecite{Lam93}, transport properties are determined by
$$ P^{\alpha,\beta}_{i,j}(\varepsilon,\hat H)=
\sum_{\nu,\nu^\prime}\left|
S_{(i,\nu),(j,\nu^\prime)}^{\alpha,\beta}
(\varepsilon,\hat H)
\right|^2, $$
which is referred to as either a reflectance $R$ ($i=j$) or a
transmittance $T$ ($i\ne j$) from quasi-particles of type $\beta$
in lead $j$ to quasi-particles of type $\alpha$ in lead $i$.
After normalization to unity with the number of open channels $N_{\rm ch}$,
$\alpha \ne\beta$, $ P^{\alpha,\beta}_{i,j}(E,H)$ defines
the Andreev scattering probability, while for $\alpha=\beta$, it
indicates a normal scattering probability. Such normalized quantities
are equivalent to an angle-average and can be directly compared to
the corresponding classical probabilities.

In the remaining of the article we focus on the low-energy solutions of
Eq.~(\ref{eqn:BdG_Hamiltonian}) with quasiparticle energy $\varepsilon=0$,
which is appropriate for the model of perfect Andreev reflection at the
NS interface. In this case, particle and hole coefficients coincide. Hence, we
adopt the shorthand notation
$R_e \equiv R^{ee} / N_{\rm ch} = R^{hh} / N_{\rm ch}$,
$R_h \equiv R^{he} / N_{\rm ch} = R^{eh} / N_{\rm ch}$
and
$T_e \equiv T^{ee} / N_{\rm ch} = T^{hh} / N_{\rm ch}$,
$T_h \equiv T^{he} / N_{\rm ch} = T^{eh} / N_{\rm ch}$
to indicate
reflection and transmission probabilities, respectively.

Due to interference effects, quantum scattering coefficients are
rapidly oscillating functions of the Fermi energy. Therefore, in order to remove
the quantum fluctuations, we perform an energy average over
values $k_F W/\pi \in [6.2, 6.8]$, which corresponds to six open channels in the
leads, $N_{\rm ch} = {\rm Int}[k_F W/\pi]$, for each value of the magnetic
field. The remaining parameters of our simulations are as follows.
The width $W$ of the lead is $25a_m$. For the superconducting antidot,
we define the pair potential via $\Delta=\hbar v_F/2\xi_S$ by choosing
$\xi_S=8a_m$ so that the diameter $2W$ is approximately 6 times larger
than the superconducting coherence length. To define the tunnel barrier in the case
of the normal antidot, an onsite potential of $100\times \hbar^2/2ma_m^2$
is added to all lattice sites lying inside. Note that the mesh lattice constant $a_m$
need not be defined explicitly if all energies are measured in $\hbar^2/2ma_m^2$
yielding $\gamma_{ij} = 1$ for every $i,j$-pair. The above definitions are
consistent with the requirements set in Ref.~\onlinecite{FTPR05} about
lengthscales, namely, $\xi_S, \lambda_F \gg a_m$ and $\xi_S/\lambda_F > 1$.
Here, $\lambda_F$ is the Fermi wavelength.

\begin{figure}
  \epsfig{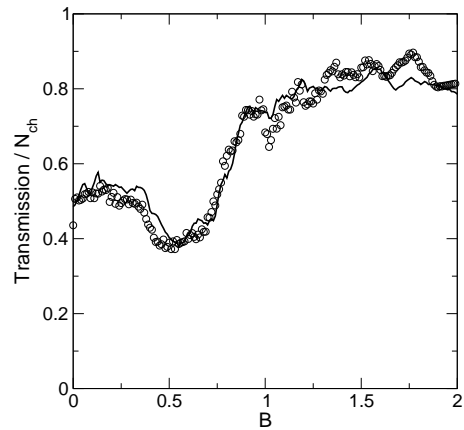}
  \caption{
  Magnetic field dependence of the
classical (solid line) and quantum (dots) transmission probability
for the normal conducting Sinai billiard of Fig.~\ref{fig:fig1}.
The magnetic field is in units of the strength $B_{0}$ for which the
cyclotron radius is equal to $W$.
}
  \label{fig:Trans_nc}
\end{figure}

First we consider the Sinai billiard with a normal antidot in the
center acting as a potential barrier. In this case coefficients with
$\alpha \ne \beta$, i.e., involving particle-to-hole conversion
(and vice-versa) are identically zero as there is no Andreev reflection at the
antidot boundary. Particles (holes) can be either normally transmitted or
reflected. In Fig.~\ref{fig:Trans_nc}, the smoothed transmission is compared
to the classical curve (Sec.~\ref{sec:S3a}) revealing the same
qualitative features. Even more remarkably, we see a very
good quantitative agreement between both curves with deviations being
within the amplitude of the small oscillations. Reflection
is just symmetric to the transmission, i.e. $R_e = 1 - T_e$,
as both the classical and the quantum calculation respect unitarity. 

\begin{figure}
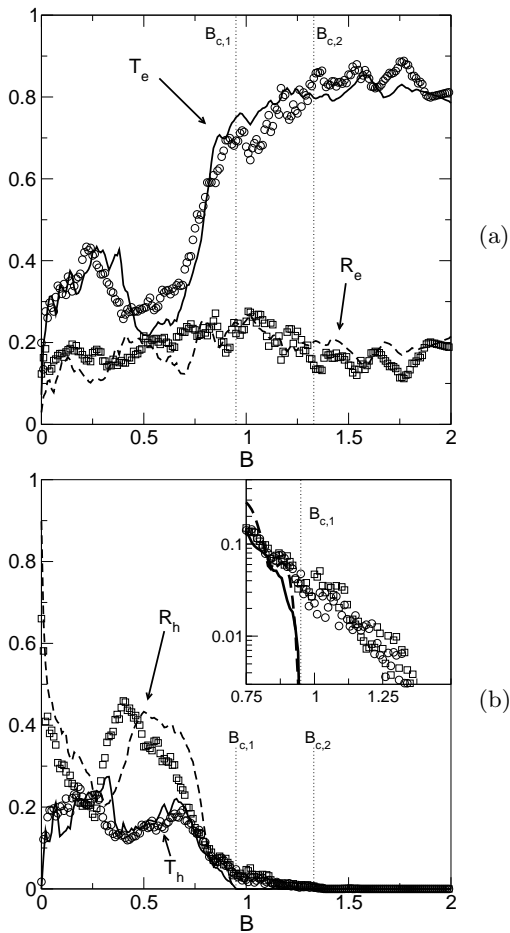

\begin{minipage}{15pc}
\centerline{\epsfig{figure=fig4a.eps,width=6cm}}  
\end{minipage}
\begin{minipage}{0.5pc}
(a)
\end{minipage}

\begin{minipage}{15pc}
\centerline{\epsfig{figure=fig4b.eps,width=6cm}}
\end{minipage}
\begin{minipage}{0.5pc}
(b)
\end{minipage}
  \caption{
Panel (a) shows the transport coefficients of particles escaping
the Andreev version of the Sinai billiard of Fig.~\ref{fig:fig1}
without particle-to-hole conversion. Panel (b) shows the Andreev
reflection and transmission probabilities. In both panels the
solid (dashed) line is the classical result of transmission
(reflection) and the circles (squares) show the quantum
transmission (reflection) coefficients normalized by $N_{\rm ch}$.
The field strength is given in units of $B_{0}$. 
Inset: semilogarithmic blow up for $0.75 \leq B / B_{0} \leq 1.25$.
}
  \label{fig:Trans_sc}
\end{figure}

Second we consider the case where the central antidot becomes superconducting.
Andreev reflection now gives rise to non-zero $R_h$ and $T_h$
coefficients as shown in panel (b) of Fig.~\ref{fig:Trans_sc}.
Upon comparison of the quantum results with the classical curves,
we see again that they agree very nicely. The results are summarized
in Fig.~\ref{fig:Trans_sc} with vertical lines indicating two distinct
values of the magnetic field, $B_{c,1}$ and $B_{c,2}$, that are related
to different qualitative features in the classical dynamics.
The largest differences occur for the particle-to-hole coefficients at the
first critical field $B_{c,1}$. There, the classical transmission and
reflection vanish abruptly, whereas the averaged quantum mechanical coefficients
decay exponentially (see inset of Fig.~\ref{fig:Trans_sc}).
However, we leave the analysis of such effects as well as the overall non-monotonic
behavior with respect to the magnetic field for the next section for a discussion
under the prism of the properties of the classical trajectories.

\begin{figure}
  \epsfig{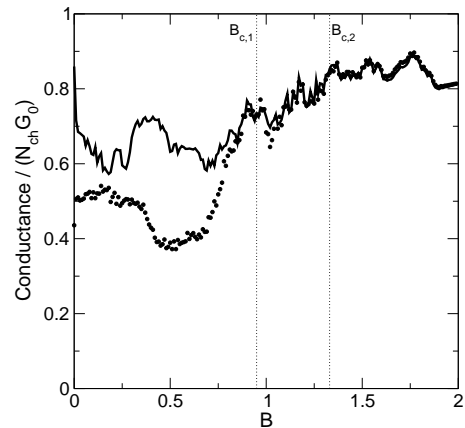}
  \caption{
Magnetoconductance for the normal (dots) and the Andreev (solid line)
Sinai-shaped billiard of Fig.~\ref{fig:fig1}  (in units of the number of
open channels, $N_{\rm ch}$, times the conductance quantum $G_0 \equiv 2e^2/h$).
The field strength is in units of $B_{0}$.
}
  \label{fig:Cond}
\end{figure}

To conclude this section, we would like to show how the conductance,
as an experimentally accessible quantity, changes when the antidot is
made superconducting. In Fig.~\ref{fig:Cond}, the magnetoconductance of a
normal (dots) and for a superconducting (solid line) antidot is plotted.
In the normal case the linear-response low-temperature conductance is
simply proportional to the transmission $T_e$, according to Landauer's
formula $G_N=(2e^2/h)\;T_e$. Lambert {\it et al}~\cite{Lam93} have worked out
generalizations for systems including superconducting islands or leads.
For the Andreev version of the Sinai billiard system of Fig.~\ref{fig:fig1},
the conductance is given by $G_S=(2e^2/h)\;(T_e+R_h)$. Overall, we see that
in the presence of Andreev reflection the conductance of the system is larger
than in the normal conducting case for magnetic fields $B<B_{c,1}$.
For larger fields the particle-to-hole
coefficients vanish and the conductances for both cases almost coincide.

\section{Classical dynamics} 
\label{sec:S3}

\subsection{General features}

In this section we study the classical dynamics of the incoming particles
(we focus on electrons but the same arguments apply to incoming holes)
for each of the two antidot structures described above.

The general form of the Hamiltonian describing the dynamics of charged particles
inside the cavity reads
\begin{equation}
H=\frac{1}{2 m^*_\alpha}(\vec{p}-q_\alpha \vec{A})^2.
\label{eq:hamiltonian}
\end{equation}
The index $\alpha$ is used to describe the possibility that the propagating
particles are either electrons ($e$) or holes ($h$). This generalization is
necessary for a correct description of the dynamics in the
setup with the superconducting antidot. The canonical momentum vector is
$\vec{p}=(p_x,p_z)= m^*_\alpha \vec{v} + q_\alpha \vec{A}$ where $\vec{v}$ is the
mechanical velocity, the corresponding position vector being $\vec{r}=(x,z)$.
Charge conservation yields $m^*_h=-m^*_e$ for the effective masses and
$q_h=-q_e$ for the electric charge.
The main property which distinguishes the two cases, i.e.,
normal/superconducting antidot, is the interaction of the charged particle
with the scattering disk. The latter is captured by the elementary processes
illustrated in Fig.~\ref{fig:fig2}, namely, specular reflection (SR) versus
the Andreev reflection (AR).

In what follows, we calculate the electronic transport properties
by analyzing the ballistic propagation and escape of 
classical 
particles injected into the billiard via the opening pipe-like channels
(see Fig.~\ref{fig:fig1}). The initial conditions for incoming
electrons are determined by the phase-space density
\begin{equation}
\begin{split}
\rho_o&(x,z,v_x,v_z) = \frac{1}{2 m^*_e v W}\delta(x+\frac{L}{2})\times \\
 & \left[\Theta(z+\frac{W}{2})-\Theta(z-\frac{W}{2})\right]
\delta(m^*_e(v-v_F))\cos \theta, \\
\label{eq:density}
\end{split}
\end{equation}
where $\theta \in [-\frac{\pi}{2},\frac{\pi}{2}]$ is the angle of the initial
electron momentum with the $x$-axis and $v_F=\sqrt{2 E_F/m^*_e}$
and the coordinate origin is assumed at the center of the cavity.

The trajectories of the charged particles in the billiard consist of segments
of circles with cyclotron radius $r=m^*_\alpha v/ (-q_e B)$
(with $v=\sqrt{v_x^2+v_z^2}$). At non-vanishing external field the classical dynamics
of both the normal and Andreev billiards is characterized by a mixed
phase space of co-existing regular and chaotic regions. 
At $B=0$ the superconducting antidot leads to an integrable dynamics,
(since trajectories are precisely retraced after retroreflection),
while the corresponding normal device possesses a mixed phase space.
It is convenient to write the dynamics (collisions with the walls and the antidot)
explicitly in the form of a discrete map. As the magnitude of the
velocity remains constant in time, a simple
parameterization of the dynamics is given by determining the position $(x_n,z_n)$
of the $n$-th collision with the boundary and the angle $\theta_n$ of the
velocity vector with respect to the normal of the boundary at the collision
point taken after the collision. Here, the term boundary refers to the walls 1
throughout 4 and the circumference of the antidot (see Fig.\ 1).

There are three families of periodic orbits each forming a continuous set
that occur in the classical dynamics and phase space of the
{\it{closed system}}~\cite{GasDor,Kov,SilAgu},
i.e. without leads, leaving their fingerprints in the {\it{open system}} with the
attached leads. We will briefly discuss these periodic orbits in the following.
At zero field there are orbits bouncing between two opposite walls 
with velocities parallel to the normal of the corresponding walls. At finite
but weak $B$-field strength the periodic orbits form a rosette
and incorporate collisions with the antidot and the walls.
These periodic orbits are typical, i.e. dominant up to a critical field value
$B_{c,2}$. For magnetic fields above $B_{c,2}$ the cyclotron radius is so small that
no collisions with the antidot can occur and 
skipping orbits, describing the hopping of the electrons along
the billiard walls, become dominant. 
All periodic orbits possess an eigenvalue one of their stability matrix~\cite{FliSchSpo}
and all periodic orbits possess unstable directions.
We remark that the above-discussed periodic orbits of the closed billiard
are not trajectories emerging from and ending in the leads of the open billiard.
However, trajectories of particles coupled to the leads (i.e., injected and transmitted/reflected)
can come close to the periodic orbits of the open billiard thereby tracing
their properties. This way the presence of the periodic orbits reflects itself
in the transport properties.

\subsection{Sinai billiard with normal antidot}
\label{sec:S3a}

\begin{figure}
\includegraphics{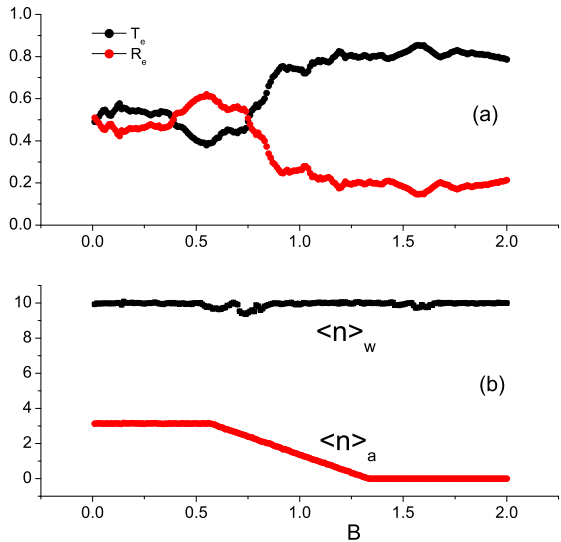}
\caption{\label{fig:fig3}
(a) Classical electron transmission, $T_e$, and reflection, $R_e$,
coefficients for the normal antidot device of Fig.~\ref{fig:fig1}
as a function of the applied magnetic field $B$.
(b) The mean number of collisions with the boundary of the square cavity
(walls 1-4), $\langle n\rangle_w$, and with the circumference of the
antidot, $\langle n\rangle_a$, as a function of $B$.
The field strength is given in units of $B_0$.}
\end{figure}

First we consider the transport of electrons through the Sinai billiard
(Fig.~\ref{fig:fig1}) with a normal antidot.
The relevant quantities determining the current flow through the device are the
transmission $T_e$ and reflection $R_e$ coefficients for electrons
defined as the percentage of the initial electrons leaving the
device from the right and left lead, respectively.
Additional quantities that are helpful for an understanding and analysis of
the system dynamics are the mean number of collisions per incoming
electron with the walls (1-4), $\langle n\rangle_w$, and with the antidot, $\langle n\rangle_a$.
We calculate these quantities by numerical simulation for different
values of the external magnetic field $B$.
It is convenient to use a dimensionless form of the classical equations
of motion by employing the scaling $x=\xi_x W$ and $z=\xi_z W$ for the spatial coordinates and
$t=\tau / \omega $ (with $\omega = B_0/m^*_e$) for the time coordinate.
The above quantities are calculated for 100 values of the magnetic
field strength varying from 0.02 to 2 using an ensemble of $10^6$ different
initial conditions distributed according to Eq.~(\ref{eq:density}) for each $B$-field value.
The  magnetic field dependence of the coefficients $T_e$ and $R_e$ is shown
in Fig.~\ref{fig:fig3}a. The obtained curves are quite irregular, 
possibly indicating the presence of fractal fluctuations in the magnetoconductance of the
system~\cite{Ketz}. 

In Fig.~\ref{fig:fig3}b, we present the parametric dependence
of the mean quantities $\langle n\rangle_w$ and $\langle n\rangle_a$.
Interestingly, the mean number of collisions with the walls remains
constant $\langle n\rangle_w \approx 10$ for almost all values of the field strength.
This means naturally that also the accumulated number of collisions of all
injected trajectories with the walls is independent of the field strength
for the whole regime considered. This number is obtained by
integrating the occupancy of the trajectories in phase space, i.e., their measure,
over all possible velocities and the boundary of the cavity (defined by the walls
1-4 including the leads). Its invariance with respect to the field strength
is of combined geometrical and dynamical origin and can be understood as follows. 
The escape probability is well approximated by $P=\mu_{\rm esc}/\mu_{T,b}$
where $\mu_{\rm esc}$ is the measure of phase space points on the left and right 
leads visited by the escaping electrons, while $\mu_{T,b}$ is the total measure
involved in the dynamics of the system along the boundary defined 
by the walls 1-4 (including the leads). The corresponding integrals can be
estimated as $\mu_{\rm esc}= 2 c_l W$ and $\mu_{T,b}= 4 c_b L$. Here, $c_l$ is the mean
phase space density on the leads and $c_b$ is the mean phase space density
on the entire boundary, both integrated over the momenta.
Due to the symmetric setup of the leads relative to both the $x$ and $z$
axis, we have $c_l=c_b=c$. Thus, $P=1/10$ and the mean number of
collisions with the wall is $\langle n\rangle_w=1/P=10$.

The behavior of $\langle n\rangle_a$ is more complicated because the dynamical
occupation of
the antidot's circumference strongly depends on the value of the external field.
One can clearly distinguish three regimes: (i) the low field region
ranging from $B \approx 0$ to $B=B_{c,1} \approx 0.55$, (ii) the intermediate
field region with $B_{c,1} < B < B_{c,2}(\approx 1.33)$, and (iii) the high field
region with $B > B_{c,2}$. All three regions are characterized by different
properties of the corresponding phase space. These are revealed
by the study of the phase space structure using Poincar\'{e} surfaces of
section (PSOS) for different values of the applied field $B$.
We employ $(x,v_x)$ sections defined by the condition $z=0$. It turns out that
all calculated surfaces of section reveal a mixed phase space.
To further quantify our analysis we calculate the relative weight
$w_c$  of those trajectories on the PSOS that exhibit collisions with the antidot.
Since collisions with the antidot are the only possibility to obtain dynamics that
is sensitive with respect to the initial conditions,
$w_c$ is also a measure for chaoticity in phase space~\cite{comm1}. We partition the
energetically allowed phase space on the $(x,v_x)$ plane into $N=10^4$ cells of
equal size and define on each cell  ${\cal{C}}_i$ the characteristic
function $h_{{\cal{C}}_i}(x,v_x)$ as:
\begin{equation}
h_{{\cal{C}}_i}=\left\{ \begin{array}{l}
1~~~ \textup{if an orbit exists with}~(x,v_x)~\in {\cal{C}}_i\\
~~~~~\textup{that hits the antidot,}  \\   
0~~~ \textup{if}~(x,v_x)~\not\in {\cal{C}}_i~\textup{holds for {\it all}}\\
~~~~~\textup{trajectories hitting the antidot.}\\ 
                       \end{array}\right.
\label{eq:charfunc}
\end{equation}
We subsequently approximate
$w_c \approx (1/N) \sum_{i=1}^N h_{{\cal{C}}_i} $.

\begin{figure}
\includegraphics{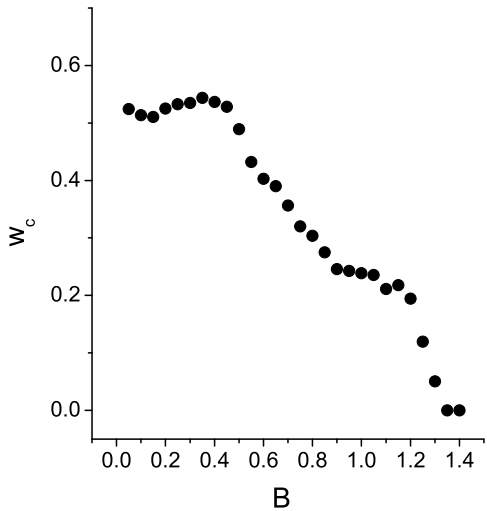}
\caption{\label{fig:fig4}
Relative weight of the chaotic part $w_c$ of the phase space 
as a function of the applied field $B$ (in units of $B_0$) for the 
normal conducting Sinai-shaped billiard of Fig.~\ref{fig:fig1}. }
\end{figure}

The function $w_c(B)$ is shown in Fig.~\ref{fig:fig4}.
The initial plateau at $w_c \approx 0.5$ shows clearly that the system is to
a large portion chaotic for low magnetic fields. This explains the fact that in
this range of fields the mean number of collisions with the antidot
$\langle n\rangle_a$ is almost constant. The degree of chaos in the phase space
of the system is large enough thereby ensuring that, with the exception of trajectories
of negligible measure, each trajectory hits the circumference of the antidot.
Hence, as evaluated in a similar fashion to $\mu_{T,b}$, the total measure of phase
space points $\mu_{C,a}$ involving the circumference is equal to $2 \pi c W$.
Following the arguments given above for $\langle  n \rangle_w$, we estimate
$\langle n\rangle_a$ as $\langle n\rangle_a/\langle n\rangle_w=
\mu_{C,a}/\mu_{T,b}=2 \pi c W/ ( 4 c L)$, yielding
$\langle n\rangle_a \approx \pi$, which is in very good agreement with
Fig.~\ref{fig:fig3}b.
In the intermediate field region the weight of the chaotic
trajectories decreases almost linearly and becomes vanishingly small in the high
field region. The linear decrease of $w_c$ 
leads to a linear decrease of $\langle n\rangle_a$ for this range of
magnetic fields. Above $B_{c,2}$ the cyclotron radius of the electron trajectories
is so small ($r < L/4-W/2$) that no collision with the antidot
is possible, i.e., $\langle n\rangle_a=0$. 

We can now understand the non-monotonic behavior of the functions
$T_e(B)$, $R_e(B)$ by considering the representative trajectory dynamics for
various $B$-values. According to Fig.~\ref{fig:fig3}a the low field region
possesses two subregions: (i) $0< B < 0.4$ and (ii) $0.4 < B < 0.55$.
Similarly the intermediate field region can be divided to:
(i) $0.55 < B < 0.75$ and (ii) $0.75 < B < 1.33$.
Note that the points defining the magnetic field windows,
$B \approx 0.4, 0.55, 0.75$ and $1.33$,
also mark qualitative changes in the functions
$\langle n\rangle_a(B)$ and $w_c(B)$.

At very low fields, $B\rightarrow 0$,
we observe that $R_e$ is slightly larger than $T_e$.
This owes to many trajectories exhibiting only one collision with the
antidot and reflected directly back to the lead 
from where they came.
The typical configuration consists of an incoming electron moving almost
as a free particle, hitting the antidot once, and escaping from the billiard to the
left lead (electron reflection). Otherwise, in
the low field region (i) the main process is the transmission of electrons. 
As the magnetic field increases, the reflection angle at the circumference of the
antidot increases too and the incoming electron, after hitting the antidot,
suffers two or more collisions with some of the walls 1,2,3 or 4 before escaping to
the right lead (electron transmission). 
This mechanism, along with a significant amount of electrons that are initially
emitted with a larger angle ($\vert\theta\vert > \tan^{-1}(2/5)$), hitting
directly the upper or lower wall of the cavity (or even its right wall)
 suffering specular reflection and exiting to the right lead
of the device, establishes electron-transmission as the main process in the low
field region. However, it is evident from Fig.~\ref{fig:fig3}a that the
difference between $T_e$ and $R_e$ is quite small.
There are complex trajectories with more than 10 collisions with the walls
and 5-8 collisions with the antidot possessing a finite measure in phase space.
These give a non-vanishing contribution to $R_e$ thereby maintaining its mean
value around $0.45$. In fact $T_e$ and $R_e$ fluctuate insignificantly around
their mean values ($0.55$ for $T_e$ and $0.45$ for $R_e$). 
As the fields increases above $B=0.4$, trajectories with a larger number of
collisions with the walls may become statistically more important
but as long as $B < 0.55$ the trajectories with several collisions with the
antidot have still a significant measure
(see Figs.~\ref{fig:fig4} and~\ref{fig:fig3}b, respectively). Overall,
for $0.4 < B < 0.55$, the main process is the reflection of electrons
yielding a large difference between $T_e$ and $R_e$. Typical trajectories
have one or two collisions with the walls and a single collision with the antidot.
The incoming electron hits the antidot once, is specularly reflected and escapes from
the billiard to the left lead, after suffering one more collision with the wall 4.

At intermediate fields, we observe an almost monotonic decrease of
$R_e$ and an increase of $T_e$ owing to the combined decrease of
$\langle n\rangle_a(B)$ and $w_c(B)$. The small plateau feature in
Fig.~\ref{fig:fig4} around $w_c=0.6$ is also reflected in the change of
slope in the transport probabilities within subregion (i). At its upper limit,
$B=0.75$, the two coefficients become equal, $T_e=R_e$. Most trajectories
have $5-8$ collisions with the walls and $1-4$ collisions with the antidot.
In window (ii) of the intermediate-field region with $B > 0.75$,
the process of transmission is much stronger than the process of reflection.
Most trajectories have a few ($\approx 5$) collisions with the walls 
and no collision with the antidot. A typical trajectory of the incoming electron,
due to the small cyclotron radius, suffers one collision with wall 1,
three collisions with wall 4, one collision with wall 3
(5 collisions with the walls in total) and then escapes from the cavity through
the right lead (electron-transmission).
There is also the case in which the incoming electron misses
the right lead of the device, and after suffering many collisions with the
walls finally escapes to the left lead, contributing to the process of reflection.
The same scenario is valid also for the high field region.

\subsection{Sinai billiard with superconducting antidot}
\label{sec:S3b}

Compared to the dynamics of the cavity with the normal antidot, the
Andreev Sinai billiard with the central superconducting
disc exhibits basic differences due to the occurrence of trajectories which suffer
Andreev reflection, instead of specular reflection, at the circumference of the
antidot. First, the complete description of the transport properties of the system
requires the introduction of two additional coefficients
describing electrons that escape as holes either to the left or the right lead;
$R_h$ (reflection) and $T_h$ (transmission), respectively.

\begin{figure}
\includegraphics{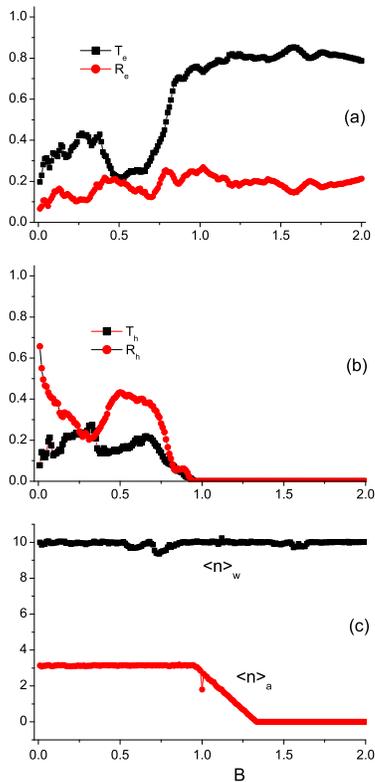}
\caption{\label{fig:fig5}
Classical results for the billiard of Fig.~\ref{fig:fig1} with a superconducting disc
in the center, as a function of the applied magnetic field $B$ 
(in units of $B_0$).
(a) Classical electron transmission, $T_e$, and reflection ($R_e$) coefficients.
(b) Classical electron-to-hole (Andreev) transmission, $T_h$, and
reflection, $R_h$, probabilities.
(c)  mean number of collisions, $\langle n\rangle_w$, with the boundary of the square cavity
(walls 1-4) and with the circumference of the antidot,
$\langle n\rangle_a$. 
}
\end{figure}

In Figs.~\ref{fig:fig5}a and~\ref{fig:fig5}b, all probabilities
are plotted as a function of the applied field $B$.
In Fig.~\ref{fig:fig5}c, we present the quantities $\langle n\rangle_w(B)$
and $\langle n\rangle_a(B)$ following the definition of Sec.~\ref{sec:S3a}.
$T_\alpha$ and  $R_\alpha$ (with $\alpha=e,h$) exhibit
irregular fluctuations as a function of $B$,
similar to those obtained for the normal billiard.
The function $\langle n\rangle_w$ is almost identical to the corresponding
function obtained for the normal case. Qualitatively, the mean number $\langle n\rangle_a$
of collisions with the superconducting antidot is also 
similar to that in Fig.~\ref{fig:fig3}b.
However, while $B_{c,2}$ is remaining the same,
the value of the critical field $B_{c,1}$ is shifted to the larger
value $B_{c,1} \approx 0.95$. The former should be expected since it does not
involve any collisions with the superconducting disc.

\begin{figure}
\includegraphics{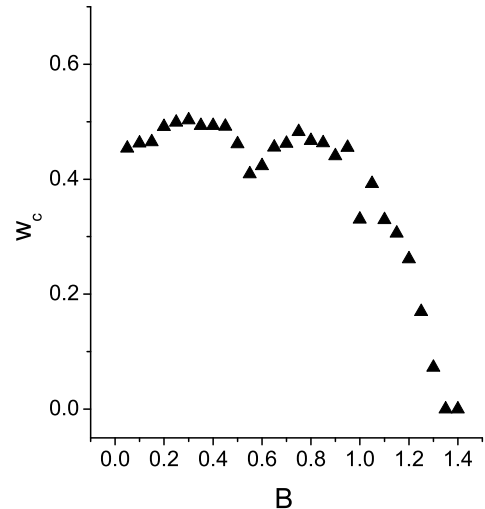}
\caption{\label{fig:fig6} Relative weight of the chaotic part of the phase space $w_c$
as a function of the applied field $B$ (in units of $B_0$) for the Andreev version
of the billiard in Fig.~\ref{fig:fig1}. 
}
\end{figure}

The difference in $B_{c,1}$ is explained by calculating the relative weight of the
chaotic trajectories as in Sec.~\ref{sec:S3a}. The result
is shown in Fig.~\ref{fig:fig6}. For magnetic fields up to $B \approx 0.95$ a
large part of the phase space of the system is chaotic ensuring the equal mean
phase space density on the boundary of the square cavity and the
circumference of the antidot. For $B > 0.95$ an almost linear decrease of $w_c$
leads to a corresponding linear decrease of $\langle n\rangle_a$. At
$B \approx 1.33$ the chaotic part of the phase space vanishes due 
to the fact that no collisions with the defocusing perimeter of the antidot
are possible. An additional peculiarity of the superconducting device
appears for the intermediate field region, $0.95 < B < 1.33$: {\em All}
possible trajectories possess an even number of collisions with the antidot,
yielding a vanishing transmission and reflection of holes, $T_h=R_h=0$.
This interesting feature is related to the generic properties of Andreev reflection
and is analyzed in Appendix A.

Let us now consider the transport coefficients in more detail.
Following the qualitative features in Figs.~\ref{fig:fig5}a and~\ref{fig:fig5}b,
we divide the low field region into four windows:
(i) $0 < B < 0.3$, (ii) $0.3 < B < 0.5$, (iii) $0.5 < B < 0.7$ and
(iv) $0.7 <B < 0.95$. Note that different regimes roughly coincide
with different qualitative features of the functions $\langle n\rangle_a(B)$ and $w_c(B)$,
as indicated previously.

In region (i), due to the large cyclotron radius
(almost vanishing curvature) most trajectories have a single
collision with the antidot. In fact, at very small fields, $B \sim 0.01$,
there is practically no collision with the walls. The typical process
consists of an incoming electron moving almost on a straight line hitting
the antidot once, 
being converted into a hole which nearly retraces the electron path
moving towards the left lead and finally leaving the device. 
Therefore, $R_h$ in this region is larger
than $T_h$, $T_e$ and $R_e$. There is however a significant amount of
electrons emitted initially with a large enough angle,
$\vert\theta\vert > \tan^{-1}(2/5)$, which hit directly the upper or lower wall
of the device suffering normal reflection and then exiting to the right lead.
This process gives a finite electron transmission coefficient,
yielding a significant contribution to $T_e$. For even larger emission
angles, there is a second but less significant set of trajectories
that exhibit specular reflections at the walls 2, 3, and 4. Particles along
these paths escape finally from the cavity through the left opening, overall
leading to  very small values of $R_e$.
With increasing $B$, the curvature of the trajectories
also increases so that an additional collision with the wall takes place. The
incoming electron hits again the antidot, becoming a hole and as the curvature
is increased the hole cannot escape from the narrow left
lead, subsequently hitting wall 1, being specularly reflected and leaving the
device from the right lead. Therefore, $T_h$ is increased at the cost of
$R_h$.

In region (ii), the curvature of the trajectories increases further. A typical 
trajectory for an incoming electron, after being Andreev reflected at the antidot, hits the wall 4 and
escapes to the left lead after specular reflection. As a result $R_h$ increases
in this region. We also observe a decrease of $T_e$
owing to the increased curvature of the trajectories making it difficult for the incoming
electrons to avoid the collision with the antidot and it is therefore harder to encounter the outgoing right lead.
In region (iii), there is no significant variation of the observables
$T_{e,h}$ and $R_{e,h}$.

As the intermediate field regime defined by the linear decrease in
$\langle n\rangle_a$ is approached,
in (iv) we observe a  sudden decrease of both $T_h$ and $R_h$
with a simultaneous increase of, predominantly, $T_e$.
Although the measure $w_c$ of trajectories with several collisions with the antidot is
non-zero, thereby, contributing to the chaotic part of the phase space,
the predominant part of the trajectories do not experience Andreev reflection.
In the regime with $B_{c,1} \sim 0.95$, we encounter $T_h=R_h=0$ due to
the fact the every trajectory has an even number of collisions with the antidot (see Appendix A).
The predominant part of the orbits have 5 to 7 collisions with the walls 1,3 and 4
thereby hopping along the boundary of the square cavity. Finally, the high-field region
is characterized by $T_h=R_h=0$, and $\langle n\rangle_a=0$ as in the case
of the normal antidot.

\section{Summary and outlook}
\label{sec:S4}
Performing simulations of the classical and quantum dynamics
of low-energy quasiparticles, we showed that a purely classical analysis
may be used as an interpretation tool for the average transport properties of
generic normal and Andreev billiards. In particular, the parametric dependence 
on the strength of a perpendicular magnetic field B was studied.
As the strength increases, this dependence of the classical trajectories
on the applied magnetic field drives the classical dynamics from mixed to regular
for both types of billiards. The latter is grossly reflected in the non-monotonic
behavior of the magnetoconductance at intermediate fields.

Owing to the increasing trajectory bending, a slight increase of the conductance $G_N$
at small fields of the normal billiard is followed by a significant valley whose minimum
defines the passage to intermediate strengths. Scattering of particles with the
Sinai-billiard disc starts reducing, also triggering the relative weight $w_c$ of the
chaotic part of phase space to shrink. Around $B_0$, which corresponds to
a cyclotron radius equal the disc radius, skipping orbits settle in and transport
properties converge towards the high-B limit.

Turning on the superconductivity at the
Sinai-billiard disc results in the interplay of the bending of the trajectories and
the occurring particle-to-hole conversion. The magnetic field drives the integrable correlated
motion of particles and holes into a mixed dynamics regime evident
by an initial tendency of $G_S(B)$ towards $G_N(B) \sim 0.5 \times (2e^2/h)$
at small fields, typical for systems with phase space having a relative large
chaotic part. Compared to the normal case, increasing $B$ Andreev reflection counteracts to
the reduction of $w_c$, which occurs eventually but at higher field strengths.
Hence, realizing the Andreev billiard leads to qualitatively different behavior.
Rather than a magnetoconductance dip, we note a reentrance effect
of the conductance towards its initial higher value.

Our classical calculations provide not only a qualitative rationalization of
the observed properties of the exact quantum mechanical scattering
coefficients and of the corresponding magnetoconductance spectrum but also allow
us to make quantitative predictions, as evidenced by the remarkable agreement
between the classical and quantum values. This is ideal for the designing of
experimental setups and a simple analysis of the results,
since classical simulations are much less time consuming.

In the present paper we studied the (energy) averaged transport properties,
i.e., removing the quantum fluctuations. Yet the averaged quantum results
contain weak localization effects at zero and small magnetic fields.
These quantum corrections to the averaged transmission are of order one (more precisely -1/4 
for chaotic ballistic systems) compared to the classical contribution  
which is proportional to $N_{\rm ch}$. More specifically, for the 
transmission per channel for the normal conducting Sinai billiard with $N_{\rm ch} =4$ 
the negative quantum correction at zero field is expected to be below 0.1,
in line with the numerical results depicted in Fig.~\ref{fig:Trans_nc}. It
is indeed possible to extend the existing semiclassical theory for ballistic weak
localization~\cite{Jalab,RS02} to averaged quantum transport through Andreev 
billiards~\cite{AL04}. However, our focus was on the finite $B$-field range where weak
localization effects do not exist. Rather, conductance fluctuations in this regime encode
additional quantum information, and previous results in closed and open billiards
(either with a normal or a superconducting lead) indicate that such fluctuations
are interwined with the underlying classical properties. Therefore, we envisage
that our study could be further developed and utilized both theoretically and
experimentally in future investigations.

\begin{acknowledgments}
GF acknowledges funding support by the Science Foundation Ireland.
AL and MS were supported by the {\em Deutsche Forschungsgemeinschaft} within
the research school GRK 638.
\end{acknowledgments}

\begin{appendix}
\section{Derivation of the first critical field}

\begin{figure} [t]
  \epsfig{file=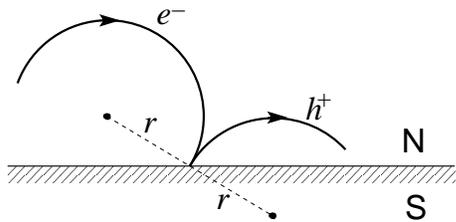, width=6cm}
  \caption{Andreev reflected orbit. The mapping from one center to the following
  one is a point reflection.}
  \label{fig:Aref}
\end{figure}

Numerical results show that for magnetic fields $B>B_{c,1}$ there are only 
trajectories with an even number of collisions with the superconducting antidot, yielding
vanishing transport coefficients $T_h$ and $R_h$. In order to derive the first
critical field it is helpful to consider the mapping of the guiding centers of the 
trajectory arcs in the presence of Andreev reflection. 
Fig.~\ref{fig:Aref} shows two segments of
an orbit right before and right after a collision with the NS-interface.
The center of the second arc can be constructed from the center of
the first one via point reflection at the collision point at the NS-interface.

\begin{figure} [t]
  \epsfig{file=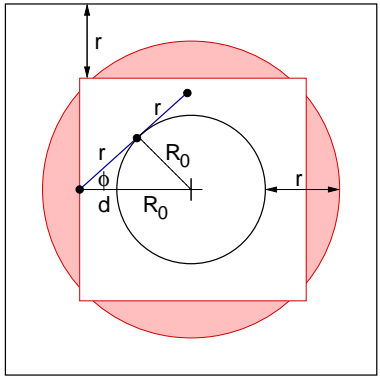, width=5cm}
  \caption{The shaded area shows all possible locations for centers of orbits
  that connect the outer walls with the antidot.}
  \label{fig:sinai}
\end{figure}

Consider one segment of an orbit with radius $r$ reaching from the outer wall to
the antidot. The distance between the center of this arc and the outer wall has
to be less than $r$ and the same holds for the distance between the center and the antidot.
This means that the center has to be inside the shaded area
shown in Fig.~\ref{fig:sinai}. Consider an orbit that hits the superconductor
only once. Such an orbit has to reach from the outer wall to the antidot and
back to the wall. So the centers before and after the Andreev reflection have to
be located inside the shaded area. This condition is easiest to fulfill for an
orbit that has its center at one midpoint of the inner quadratic boundary, as
shown in Fig.~\ref{fig:sinai}. The question is if for a certain radius the
center after the Andreev reflection can no longer be mapped into the shaded
region. 

\begin{figure} [t]
  \epsfig{file=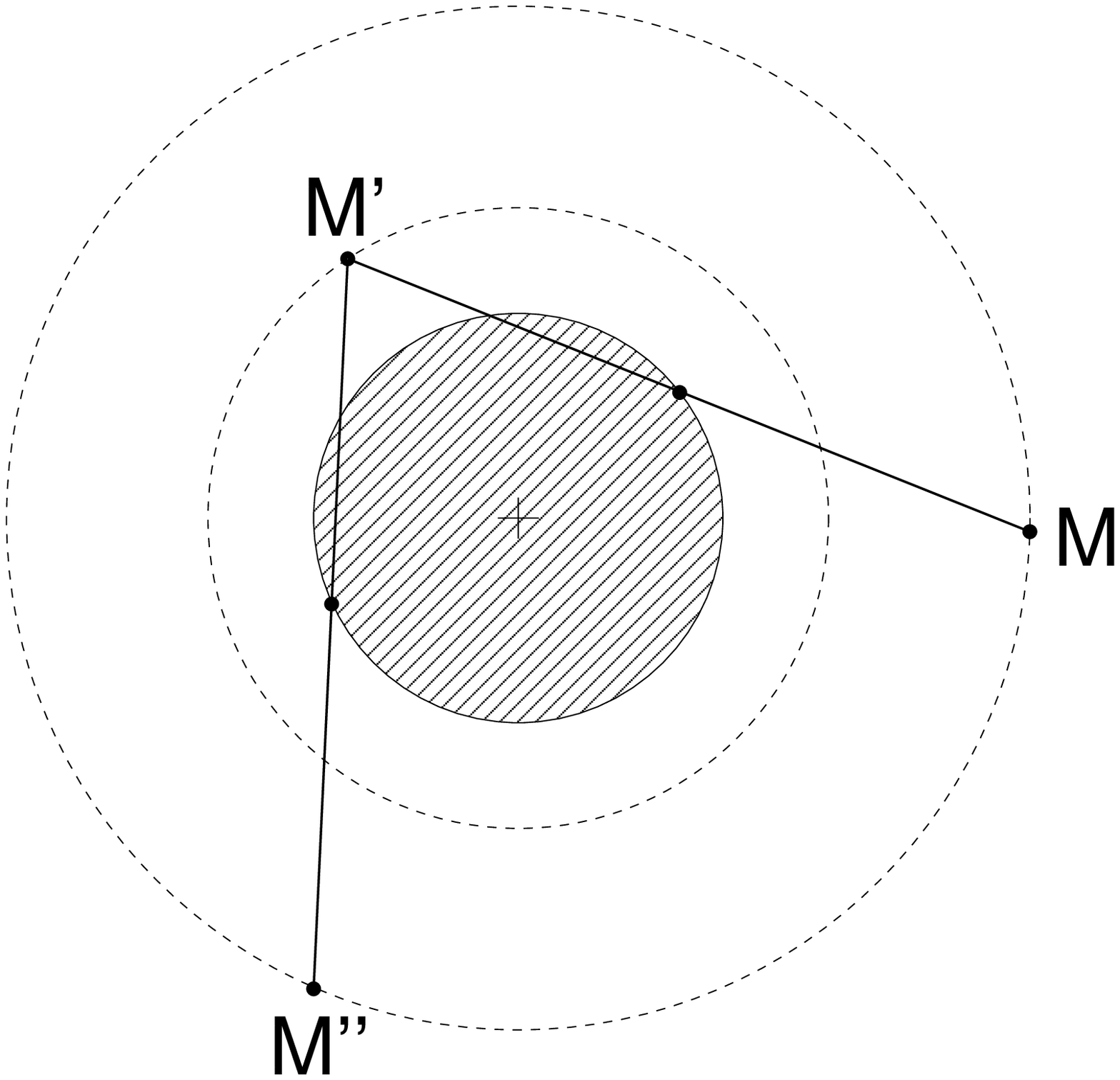, width=4cm}
  \caption{$M$ is the center of the first arc before the Andreev reflection.
  After one Andreev reflection it is mapped to $M'$, after another one it is
  mapped to $M''$, which has the same distance from the center of the antidot as
  $M$.}
  \label{fig:rot}
\end{figure}

The distance $d$ between the antidot and the inner square is $d=L/2-r-R_0$,
where L is the side length of the square cavity and $R_0$ is the radius of the
antidot. Applying the cosine-theorem to the triangle shown in Fig.~\ref{fig:sinai}
we get $R_0^2 = r^2+(R_0+d)^2-2r(R_0+d)\cos\phi$. So the angle
$\phi$ can be written as
\begin{equation}
  \cos\phi = \frac{L^2-4R_0^2+8r^2-4Lr}{4r(L-2R)}.
\end{equation}
The distance between the final point and the central horizontal line is
$y=2r\sin\phi=2r\sqrt{1-\cos^2\phi}$. In order to get only one single Andreev
reflection, the final point has to be inside the shaded region, therefore we have
to claim $y>R_0+d$, which means
\begin{equation}
  2r\sqrt{1-\cos^2\phi}-\frac{L}{2}+r>0.
\end{equation}
Solving this inequality for $L=5W$ and $R_0=W$ we find a critical radius
$r_c=1.0505W$, which corresponds via $r/W = B_0/B$ to a critical field of
\begin{equation}
  B_{c,1} = 0.9519B_0 \, .
\end{equation}

Up to now we have only shown that for a magnetic field $B>B_{c,1}$ the particles
hit the superconductor at least twice consecutively. But now it is easy to see
that the number of Andreev reflections is indeed even. Two Andreev reflections
in a row correspond to a rotation of the center around the center of the
antidot, as shown in Fig.~\ref{fig:rot}. After an even number of Andreev
reflections the center of the arc is always located on the outer dashed circle,
which has a radius greater than $R_0+d$. After an odd number of reflections at
the superconductor it is on the inner dashed circle with a radius smaller than
$R_0+d$. Therefore the particle can only `escape' the superconductor after an
even number of collisions, which explains the fact.

\end{appendix}


\begin{thebibliography}{----}
\bibitem{QuaCh}
K.-F. Berggren and S. {\AA}berg (eds), {\it QUANTUM CHAOS Y2K: Proceedings
of Nobel Symposium 116}, (World Scientific, 2000).

\bibitem{Alha}
Y. Alhassid, Rev. Mod. Phys. {\bf 72}, 895 (2000).
 
\bibitem{Jalab}
R. A. Jalabert in {\it New Directions in Quantum Chaos}, G. Casati, I. Guarneri,
and U. Smilansky (eds) (IOS Press, Amsterdam, 2000).

\bibitem{Ri00}
 K. Richter, {\em Springer Tracts in Modern Physics} {\bf 161}, 
         (Springer, Berlin, 2000).

\bibitem{KosMasGol}
I. Kosztin, D. L. Maslov, and P. M. Goldbart, Phys. Rev. Lett. {\bf 75}, 1735 (1995).

\bibitem{S-billiards}
For a recent review see, e.g., C. W. J. Beenakker, cond-mat/0406018.

\bibitem{EPL02ETW}
J. Eroms, M. Tolkiehn, D. Weiss, U. R\"{o}ssler, J. DeBoeck, and S. Borghs,
Europhys. Lett. {\bf 58}, 569 (2002).

\bibitem{And64}
A. F. Andreev, JETP {\bf 19}, 1228 (1964)

\bibitem{MBFC96}
J. Melsen, P. Brouwer, K. Frahm, and C. Beenakker, Europhys. Lett. {\bf 35}, 7 (1996).

\bibitem{PRB00CBA}
A. A. Clerk, P. W. Brouwer, and V. Ambegaokar, Phys. Rev. B  {\bf 62},
10226 (2000).

\bibitem{EPJB01ILV}
W. Ihra, M. Leadbeater, J. L. Vega, and K. Richter,
Eur. Phys. J. B {\bf 21}, 425 (2001).

\bibitem{PRL99SB}
H. Schomerus and C. W. J. Beenakker, Phys. Rev. Lett. {\bf 82}, 2951 (1999).

\bibitem{TA01}
D. Taras-Semchuk and A. Altland, Phys. Rev. B {\bf 64}, 014512 (2001).

\bibitem{PRB04CPP}
J. Cserti, P. Polin{\'a}k, G. Palla, U. Z\"{u}licke, and C. J. Lambert,
Phys. Rev. B {\bf 69}, 134514 (2004).

\bibitem{FTPR05}
G. Fagas, G. Tkachov, A. Pfund, and K. Richter,
Phys. Rev. B {\bf 71}, 224510 (2005).

\bibitem{Fer97}
D. K. Ferry and S. M. Goodnick, {\it Transport in Nanostructures}
(Cambridge University Press, 1997)

\bibitem{Ket99}
J. B. Ketterson and S. N. Song, {\it Superconductivity} (Cambridge
University Press, 1999).

\bibitem{Tad99}
F. Taddei, S. Sanvito, J. H. Jefferson, and C. J. Lambert, Phys.
Rev. Lett. {\bf 82}, 4938 (1999).

\bibitem{Lam93}
C. J. Lambert, V. C. Hui, and S. J. Robinson, J. Phys.: Cond. Mat.
{\bf 5}, 4187 (1993).

\bibitem{GasDor}
P. Gaspard and J. R. Dorfman, Phys. Rev. E {\bf 52}, 3525 (1995).

\bibitem{Kov}
Z. Kov\'{a}cs, Phys. Rep. {\bf 290}, 49 (1997).

\bibitem{SilAgu}
L. G. G. V. Dias Da Silva and M. A. M. de Aguiar, Eur. Phys. J. B {\bf 16}, 719 (2000).

\bibitem{FliSchSpo}
M. Fliesser, G. J. O. Schmidt, and H. Spohn, Phys. Rev. E {\bf 53}, 5690 (1996).

\bibitem{Ketz}
A. S. Sachrajda {\it et al.}, Phys. Rev. Lett. {\bf 80}, 1948 (1998).

\bibitem{comm1}
In a strict sense, there is no chaos since all trajectories remain in the cavity
for a finite time only.

\bibitem{RS02}
K. Richter and M. Sieber, Phys. Rev. Lett. {\bf 89}, 206801 (2002).

\bibitem{AL04}
For a chaotic Andreev billiard one finds for the weak localization correction
to the averaged conductance $-(2e^2/h)(1 + W_{\rm sc}/W)/(2+ W_{\rm sc}/W)^2$, where 
$W_{\rm sc}$ is the size of the interface with the superconductor
and $W$ is the width of the leads attached, see:
A. Lassl, diploma thesis, Universit\"at Regensburg, unpublished, 2003.
\end{thebibliography}
\end{document}